\documentclass{article}

\usepackage{amsmath}
\usepackage{amsfonts}
\usepackage{amssymb}
\usepackage{authblk}

\usepackage{caption}
\usepackage{subcaption}
\usepackage{graphicx}
\usepackage[all]{nowidow}
\usepackage[utf8]{inputenc}
\usepackage{hyperref}

\begin{document}
\date{\vspace{-5ex}}
\title{Statistics of paths on graphs with two heavy roots}
\author{Z.D. Matyushina}
\affil{Moscow Institute of Physics and Technology, 141700 Dolgoprudny, Moscow , Russia Federation}

\maketitle
\begin{abstract}

The paper considers the behaviour of the number of paths of length $N$ on graphs with two heavy roots. Such vertices can be entropic traps. Numerical analysis is carried out for graphs with different degrees of root vertices. In the symmetric case, a numerical analysis of the number of paths is performed. It is found that localisation is observed with the same ratio of vertex degrees as in the case of a single root. An analysis of the entropy forces of interaction was also carried out for two roots.

\end{abstract}

\section{Introduction}

The random walk problem studies the distribution of endpoints under a random walk, i.e.
when each next point of the path on the graph is chosen randomly in the neighbouring vertices. However, there is a problem that can be called a path counting problem, which consists in finding the number of all possible paths of length $N$.
The difference between these problems is in the different scaling of the elementary step: in the
path counting problem, all steps are included in the partition function with a weight of one, and
for random walks, the step probability depends on the degree of the vertex $p$ and is equal to
$p-1$. For heterogeneous graphs, this difference is fundamental and can lead to
"entropic" localisation\cite{Maritan1989,Ternovsky1992}.

The development of the theory of localisation in disordered systems goes back to the works of
I.M. Lifshitz \cite{Lifshitz1988,Lifshitz1968TheoryOF}. Localisation problems are often studied in polymer physics.
 For example, localisation is observed in polymer networks that are adsorbed at
at particular points in space, such as point defects. The localisation of paths on a non-homogeneous graph has an entropy character and is determined by geometric reasons. Similar problems are discussed in the literature \cite{Burda2009,Balagurov1974,Donsker1975}. 

The problem of entropy localisation on graphs of a special form has been considered in \cite{Burda2009}. In this paper, a polymer in a channel is considered as a path of $N$ steps on a graph which is $k$ rays converging at one point (Fig. 1).
\begin{figure}[h!]
\centering
\includegraphics[scale=0.5]{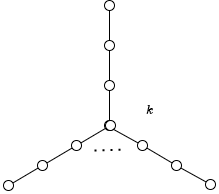}
\caption{An example of a graph with $k$ rays.}
\label{fig:Fig1.png}
\end{figure}

Therefore, for a polymer, the branching point of the channel can be an entropy trap, allowing the polymer to occupy more compact states than in the case of a straight channel without branching.
The problem to be solved in this case is to find the value of $Z_N(x)$ for such a graph. In this case, the recurrence relation is solved using generating functions. It is observed that for $k>2$, the root point becomes more entropy preferable to serve as an "entropy trap" for paths.

On a symmetric graph containing inhomogeneity (a higher degree root), path localisation is observed for the trajectory counting task, in contrast to the random walk problem. However, the phenomenon of entropy trap is not observed on homogeneous graphs \cite{Ochab2012}.

The study of the phenomenon of localisation on a wider class is presented in \cite{Nechaev2016}. The methods presented there are used in this paper to study entropy problems on trees with two heavy roots.

The analysis of the recursive relation for a tree with one heavy root, carried out in \cite{Nechaev2016}, using the generating function and the Fourier transform, allows us to find the singular values of the generating function for $\Bar{Z}_N(p,p_0)$. Thus, the asymptotic behaviour of $\Bar{Z}_N(p,p_0)$ is determined by the singularity with a smaller absolute value.
Corresponding singularities are
\begin{equation}
\sigma_1=\frac{1}{2\sqrt{p-1}},\;\;\;\sigma_2=\frac{1}{\sqrt{p}},\;\;\;\sigma_3= \frac{\sqrt{p_0-p+1}}{p_0}
\end{equation}

It can be seen that $\sigma_1>\sigma_2,\;\sigma_3$ for any $p$, but $\sigma_2=\sigma_3$ at the point $\Bar{p}_0=p^2-p$. So at this point there is a transition from a non-localised state ($p_0<\Bar{p}_0$) to a localised state ($p_0>\Bar{p}_0$).

A graph with a heavy root is an infinite (or finite) tree with one vertex of degree $p_0$ and the rest of degree $p<p_0$ (or degree 1 in the finite case). A graph with $k>1$ heavy roots is a tree in which the root vertex is connected by direct chains with k heavy roots (Fig. 2). We will consider small $k\ll p_0$.\\

\begin{figure}[h!]
\centering
\includegraphics[scale=0.4]{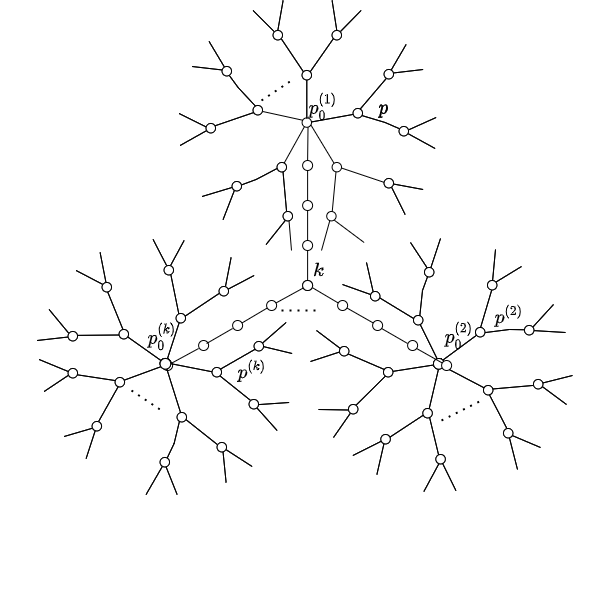}
\caption{Graph with $k$ heavy roots.}
\label{fig:Fig2.png}
\end{figure}

Let's introduce $Z_N(x)$ equals to the number of all trajectories consisting of $N$ steps and ending at a distance $x$ from the starting point. Depending on the task, we can consider different points as the starting point: the vertex of one of the heavy roots or the branching point of linear chains.
Using $Z_N(x)$, we can calculate average values:
\begin{equation}\left<x^2(N|p^{(1)}_0,p^{(1)},...,p^{(k)}_0,p^{(k)})\right>=\frac{\sum\limits_{x=0}^{\infty}x^2Z_N(x|p^{(1)}_0,p^{(1)},. ..,p^{(k)}_0,p^{(k)})}{\sum\limits_{x=0}^{\infty}Z_N(x|p^{(1)}_0,p^{(1)},...,p^{(k)}_0,p^{(k)})}
\end{equation}
In the $N\rightarrow\infty$ limit, for some systems the value of $\left<x^2(N)\right>$ does not depend on $N$\cite{Burda2009}.

On the other hand, if we consider a symmetric problem in which the degrees of all root vertices, the degrees of the branch vertices, and the distances from the starting vertex to the root vertices are the same (the starting point corresponds to the branch point of linear chains), then $Z_N(x)$ is equal to the number of ways to realise the same states. Thus we can consider the entropy of the system:
\begin{equation}
S(r)=-\sum\limits_{x=0}^\infty p(x) \ln p(x)=-\sum\limits_{x=0}^\infty \frac{Z_N(x)} {\sum\limits_{x=0}^\infty {Z_N(x)}}\ln\bigg(\frac{Z_N(x)}{\sum\limits_{x=0}^\infty {Z_N(x)}}\bigg)
\end{equation}
Where $p(x)$ is the probability of finding the second end of the chain at a point $x$ away from the start.

\section{Path counting on finite graphs with two heavy roots}

Consider an arbitrary graph $\mathfrak{G}$ with adjacency matrix $B^N_\mathfrak{G}$. Then the matrix element $\left< i\left|B^N_G\right|j \right>$ coincides with the number of paths of length $N$ from vertex $i$ to vertex $j$ starting at vertex $i$ and ending at distance $x$ from it, then

\begin{equation}Z_N^{(i)}(x)=\sum_{j:dist(i,j)=x} {\left< i\left|B^N_G\right|j \right>}\end{equation}

Therefore, if $N+x$ is even, then the asymptotic behavior of $Z_N$ for large $N$

\begin{equation}\log Z_N(x)\approx N\log\lambda_{max}(G)+o(N)\end{equation}

And $Z_N(x)=0$ if $N+1$ is odd. A graph with two heavy roots is a tree, which means that it is a bipartite graph. For such graphs there is always a pair of largest eigenvalues $\pm \lambda_{max}(G)$.\\

According to the theorem \cite{Rojo2005}, the spectral numbers of such a graph can be expressed in terms of a matrix of a special form. Thus the set of eigenvalues of the matrix $B^N_\mathfrak{G}$ coincides with the set of eigenvalues of the principal submatrices $A_k,\;k\in{\{1,\;2,\;...,\;n-r ,\;n\}}$ of a tridiagonal symmetric matrix $A^{(n)}$ with elements $a_{ij}$ defined as follows:
\begin{equation}
\begin{cases}
a_{i,i}^{(n)}=0 \\ 
a_{n,n-1}^{(n)}=a_{n-1,n}^{(n)}=\sqrt{2};\\
a_{i,i-1}^{(n)}=a_{i-1,i}^{(n)}=1; \;\;\;\;\; (i = n-r,\;..,\;n-1)\\
a_{n-r+1,n-r}^{(n)}=a_{n-r,n-r+1}^{()}=\sqrt{p_0-1}\\ 
a_{i,i-1}^{(n)}=a_{i-1,i}^{(n)}=\sqrt{p-1}; \;\;\;\;\; (i = 2,\;\;..,\;n-r-2) \end{cases}
\end{equation}

To search for eigenvalues, it is necessary to find the determinant of matrices of the form $A_k-\lambda I$. The determinant of any such matrix is determined by the recursive relation:
\begin{equation}
\det(A_k)=\lambda A_{k-1}-b_{n-1}^2A_{k-2},
\end{equation}
where $b_{n-1}$ is the $n$th element of the subdiagonal (since the matrix is symmetric, it is the same for the upper and lower subdiagonals). Although the eigenvalue relations are not solved numerically, only the largest numerical value is of interest. $\lambda$ is the largest eigenvalue if $p_k(\lambda)=det(A^{(n)}k-\lambda I)$, have the same sign for each $k$ \cite{Golub1996}.

We use the considerations of \cite{Burda2009} and take into account that for large $N$, based on the solution of the problem of finding $Z_n(x)$ on a ray and on a graph with a heavy root, $Z_n(x)$ is small for $x\leq r$. Then for an infinite graph for $x > r$ we can give the following estimate $Z_{N(x, p, p_0, n=\infty)}\geq C(p,p_0)p^N$. Similar to the case of a finite dimensional tree, there is a critical value $\Bar{p}_0$ at which the adjacency matrix has an eigenvalue greater than $p$. From the results of \cite{Nechaev2016}, an upper bound $p^2-p+1$ can be given for the critical value. On the other hand, a direct substitution of $p^2-p$ into the corresponding matrix makes it possible to verify that in this case all principal minors of the matrices are non-negative. Thus, for an infinite dimensional problem, one can expect a transition from a non-localised to a localised regime when going from the value $p^2-p$ to the value $p^2-p+1$.
Here are graphs of spectral densities in localized and non-localized modes (Fig. 3):
\begin{figure}[h!]
\begin{minipage}[h!]{0.45\linewidth}
\center{\includegraphics[width=1\linewidth]{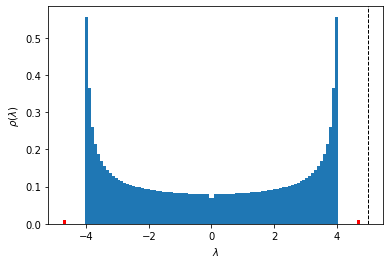} \\a)}
\end{minipage}
\hfill
\begin{minipage}[h!]{0.45\linewidth}
\center{\includegraphics[width=1\linewidth]{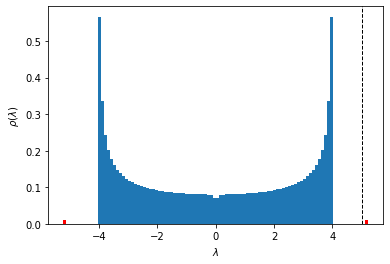} \\b)}
\end{minipage}
\caption{Spectral density plot for $p=5$ and $r=50$. a) $p_0 = 18$, non-localized mode; b) $p_0 = 22$, localized mode.}
\label{fig:Fig3.jpg}
\end{figure}

\section{Trajectory localization on an infinite symmetric graph with two heavy roots}

Of greatest interest is the case of an infinite graph, because in this case the transition to a localised state is possible. Consider an infinite graph with two heavy roots. Recursive relation for the partition function of all paths of length $N$:

\begin{equation}
\begin{cases}
Z_{N+1}(x) =(p-1)Z_N(x-1)+Z_N(x+1),\;\;\;\;\; x> r+2\\
Z_{N+1}(x) =(p_0-1)Z_N(x-1)+Z_N(x+1),\;\;\;\; x=r+1 \\
Z_{N+1}(x) =Z_N(x-1)+Z_N(x+1),\;\;\;\;\;\;\;\;\;\;\;\;\;\;\;\; 2 \leqslant x \leqslant r \\
Z_{N+1}(x) =2Z_N(x-1)+Z_N(x+1),\;\;\;\;\;\;\;\;\;\;\;\;\;\; x=1 \\
Z_{N+1}(x) =Z_N(x+1),\;\;\;\;\;\;\;\;\;\;\;\;\;\;\;\;\;\;\;\;\;\;\;\;\;\;\;\;\;\;\;\;\;\;\;\; x=0 \\
Z_{N=0}(x) = \delta_{x,1}.\\
\end{cases}
\end{equation}

We introduce a generating function:
\begin{equation}
{\mathcal W}(s,x)=\sum_{N=0}^{\infty}W_N(x)s^N\;\;\;\;\;\;\; \left( W_N(x)=\frac{1}{2\pi i}\oint {\mathcal W}(s,x)s^{-N-1}ds \right),
\end{equation}
and its Fourier transform:
\begin{equation}
\widetilde{\mathcal W}(s,q)=\sum_{x=0}^{\infty}{\mathcal W}(s,x)\sin{qx}\;\;\;\;\; \;\; \left( {\mathcal W}(s,x)=\frac{2}{\pi}\int\limits_{0}^{\pi} \widetilde{\mathcal W}(s,q))\sin {qx}dq\right).
\end{equation}
Similarly to \cite{Nechaev1987} we shift $x$, $x \xrightarrow{} x+1$ :
\begin{equation}
Z_N(x)=A^NB^x W_N(x).
\end{equation}
Where $A=B=\sqrt{p-1}$. Thus have:
\begin{equation}
\begin{cases}
W_{N+1}(x)=W_N(x-1)+W_N(x+1)+\frac{3-p}{p-1}W_N(x-1)\delta_{2,x}+ \\
\;\;\;\;\;\;\;\;\;\;\;\;\;\;\;\;\;+\frac{2-p}{p-1}W_N(x -1)\sum\limits_{i=3}^{r+1} \delta_{i,x}+\frac{p_0-p}{p-1}W_N(x-1)\delta_{r+2 ,x}\\
W_{N+1}(x)=0\\
W_{N=0}(x)=\frac{\delta_{x,1}}{\sqrt{p-1}}.
\end{cases}
\end{equation}
By performing the transformations obtain:
$$\widetilde{\mathcal W}(s,q)=\frac{1}{\sqrt{p-1}}\frac{\sin{q}}{1-2s\cos{q}}+\frac{2}{\pi}\frac{2-p}{p-1}\frac{s\cdot\sin{(2q)}}{1-2s\cos{q}}\int\limits_{0}^{\pi}\widetilde{\mathcal W}(s,q)\sin{q}\,dq+$$
\begin{equation}\;\;\;\;\;\;\;\;+\frac{2}{\pi}\frac{2-p}{p-1}\sum_{i=3}^{r+1}\frac{s\cdot\sin{(iq)}}{1-2s\cos{q}}\int\limits_{0}^{\pi}\widetilde{\mathcal W}(s,q)\sin{((i-1)q)}\,dq+\end{equation} $$\;\;\;\;\;\;\;\;+\frac{2}{\pi}\frac{p_0-p}{p-1}\frac{s\cdot\sin{((r+2)q)}}{1-2s\cos{q}}\int\limits_{0}^{\pi}\widetilde{\mathcal W}(s,q)\sin{((r+1)q)}\,dq$$
Let us denote the following integrals:
\begin{equation} I_k=\int\limits_{0}^{\pi}\widetilde{\mathcal W}(s,q)\sin{(kq)}\,dq\;\;\; \:\:\:  1\le k\le r+1. \end{equation} 
\begin{equation}
J_{n,k}=\int\limits_{0}^{\pi}\frac{s\sin{(n\cdot{q})}\sin{(k\cdot{q})}}{1-2s\cos{(r+1)q}}
\end{equation}
For $I_k$ we obtain a system of equations with the following matrix of coefficients:
\begin{equation}
\begin{pmatrix}
-1+\frac{2s}{\pi}\frac{2-p}{p-1}J_{2,1} & \frac{2s}{\pi}\frac{2-p}{p-1}J_{3,1} & ... & ... &\frac{2s}{\pi}\frac{2-p}{p-1}J_{r+1,1} & \frac{2s}{\pi}\frac{p_0-p}{p-1}J_{r+2,1}\\
... & ... & ... & ... & ... & ...\\
... & ... & ... & ... & ... & ...\\
... & ... & ... & ... & ... & ...\\
\frac{2s}{\pi}\frac{2-p}{p-1}J_{2,r+1} & \frac{2s}{\pi}\frac{2-p}{p-1}J_{3,r+2} & ... & ... &\frac{2s}{\pi}\frac{2-p}{p-1}J_{r+1,r+1} & -1+\frac{2s}{\pi}\frac{p_0-p}{p-1}J_{r+2,r+2}.\\
\end{pmatrix} 
\end{equation}
Resolving the system and applying the inverse Fourier transforms:
$${\mathcal W}(s,x)=\frac{2}{\pi}\int\limits_{0}^{\pi}\widetilde{{\mathcal W}}(s,q)\sin{(qx)}dq= \frac{2}{\pi}\frac{1}{\sqrt{p+1}}J_{1,x}+\frac{4s}{\pi^2}\frac{2-p}{p-1}J_{2,x}\frac{\vartriangle_1}{\vartriangle}+$$
\begin{equation}+\frac{4s}{\pi^2}\frac{2-p}{p-1}\sum_{i=3}^{r+1}J_{i,x}\frac{\vartriangle_{i-1}}{\vartriangle}+\frac{4s}{\pi^2}\frac{p_{0}-p}{p-1}J_{r+2,x}\frac{\vartriangle_{r+1}}{\vartriangle}\end{equation}

We perform inverse Fourier transforms:
\begin{equation}\bar{Z}_N(p,p_0)=\frac{1}{2\pi i}\oint \bar{\mathcal Z}(\sigma |p,p_0)\sigma^{- N-1}d\sigma\end{equation}
Thus, using the analytically obtained value of the integral $J_{k,x}$, we obtain a geometric progression with the same denominator as for one heavy root. Therefore, the first two singularities do not differ from the corresponding ones for one root: $\sigma_1=\frac{1}{2\sqrt{p-1}},\;\;\;\sigma_2=\frac{1}{\sqrt{ p}}$. The third singularity is not calculated for an arbitrary case, however, as a result of computer calculating, we find that the determinant of the system matrix vanishes at $\sigma=\frac{1}{p}$ between the values $p_0=p^2- p$ and $p_0=p^2-p+1$, which agrees with the estimates obtained for the finite tree. We can assume that a change in the dominant singular value leads to a change in the regime, similarly to the case with two heavy roots. This behavior is confirmed by direct iteration of the equation (8) (Fig. 5)


It is also possible to numerically analyze the case with different degree of heavy roots. If the starting point is in a subcritical root, and the distance between the points is sufficiently large, then localization will be observed only with a very significant degree of heavy root.
\begin{figure}[h!]

\hfill
\begin{minipage}[h]{0.45\linewidth}
\center{\includegraphics[width=1\linewidth]{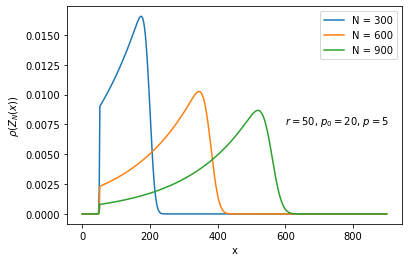} \\a)}
\end{minipage}
\hfill
\begin{minipage}[h]{0,45\linewidth}
\center{\includegraphics[width=1\linewidth]{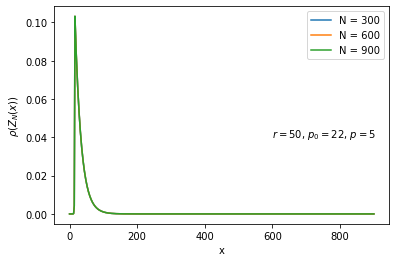} \\b)}
\end{minipage}
\caption{Graphs of the dependence of the density of the number of paths of length $x$ on $x$.
a) Roots with subcritical degree.
b) Roots with degree above critical.}
\label{fig:Fig4.jpg}
\end{figure}

Plotting the entropy on the length of the chain connecting the heavy roots (Fig. 5) allows to determine the direction of the 
entropic forces. The roots are attracted to each other, but for a sufficiently long chain connecting the roots, the force disappears and the roots cease to interact.
\begin{figure}[h!]
\centering
\includegraphics[scale=0.5]{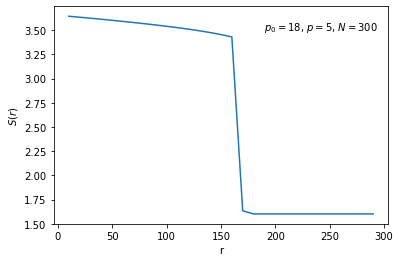}
\caption{Plot of entropy versus distance between roots for $N=300$}
\label{fig:iFig5.png}
\end{figure}
Another phenomenon that can be observed during simulation is the localization of the end of a short chain for small $N>r$, while the condition for critical $r$ coincides with the condition for the disappearance of entropy forces, and the plot of $r(N)$ turns out to be almost linear.

\section{Conclusion}
Thus, for the symmetric problem, it was found that the critical value for the degree of the heavy root lies between $p^2-p$ and $p^2-p+1$ (but since the vertex degrees take only integer values, the minimum $p_0$ at which the solution is already localised is $p^2-p+1$). Using this value to test the singular value hypothesis, it was shown that the problem does indeed enter a localised mode at such a vertex degree.

Simulation allowed to confirm the analytically obtained critical value for $p_0$ and to analyse the asymmetric case as well as other features of the statistics on such graphs. Thus, on non-symmetric graphs, the distance between the roots significantly affects the estimate for the heavy critical degree, in contrast to the symmetric case. It is also possible to observe, for some sets of $N, r, p, p_0$, a localisation at the centre of the chain connecting the graphs. In this case, for the parameters at which there is a transition from localisation at the centre of the chain to localisation at the roots (or the absence of localisation for an insufficient degree of the root), the entropy force "tightening" the roots vanishes.

The results obtained can easily be extended to trees with a large number of heavy roots $k$ (such that $k<p$).

Problems of purely geometric entropy traps have long been considered and
polymer chemistry. Applying different approaches to the problem
pure entropy localisation allows a better understanding of the problem
statistics of paths on such graphs.
\section*{Funding}

This work is supported by Basis Foundation №20-1-1-23-1.
The author thanks the supervisor A.S. Gorsky. 

\bibliographystyle{ieeetr}
\bibliography{main}
\end{document}